\documentstyle[preprint,floats,prd,aps]{revtex}

\tightenlines 
\begin{document}
%
\preprint{\vbox {\hbox{OCHA-PP-172}}}

\draft
\title{Inclusive dileptonic rare B decays with an extra generation of vector-like quarks} \author{Mohammad R. Ahmady
\footnote{Email: ahmady@phys.ocha.ac.jp}
, Makiko Nagashima\footnote{Email: g0070508@edu.cc.ocha.ac.jp} and Akio
Sugamoto\footnote{Email:sugamoto@phys.ocha.ac.jp}}
\address{Department of Physics, Ochanomizu University \\
1-1 Otsuka 2, Bunkyo-ku,Tokyo 112, Japan}

\date{May 2001}
\maketitle
\begin{abstract}
We investigate the leading effects of extending the Standard Model of electroweak interactions by an extra iso-singlet up- and down- type quark pair on various distributions and total branching ratio of the inclusive $B\to X_s\ell^+\ell^-$ ($\ell =e,\mu $) rare B decays.  The presence of the extra vector-like down quark $D$ results in the non-unitarity of the extended quark mixing matrix $V$, which in turn leads to $b\to s$ FCNC at the tree level proportional to $(V^\dagger V)_{sb}$.  On the other hand, the effective penguin and box vertex functions are sensitive to the mass of the extra iso-singlet up quark $m_U$.  The experimental upper bound on $BR(B\to X_s\mu^+\mu^-)$ is used to constrain the parameters of the model.  It is shown that the shapes of the differential branching ratio and forward-backward asymmetry distribution are very sensitive to the value of the model parameters.  We also calculate the CP aymmetry distribution of the dileptonic decay in the vector-like quark model.  It is shown that, for a typical choice of the model parameters, asymmetries up to around 10\% can be achieved for certain values of the dilepton invariant mass.  
\end{abstract}
%
\newpage

\section{Introduction}
Rare B decays, which proceed via loop effects, are important venues for observing the signals of new physics beyond the Standard Model.  Heavy exotic particles, which are not accessible to experiments at present or near future accelerators, can occur in quantum loops as virtual particles, and therefore, be detected indirectly.  B factories, CLEO III and other dedicated B experiments are expected to observe new rare B decay channels and to improve the precision of those which have already been measured.  As a result, by using experimental data on radiative rare B decays, one should be able to constrain the parameters of new physics better than before.

One simple extension of the SM is obtained by adding an extra generation of iso-singlet quarks\cite{vqm}.  There are various motivations for this model: Down-type vector-like quarks are predicted by certain grand unified theories like $E_6$ in four dimensions, which is inspired by superstring $E_8\times E_8$ model in ten dimensions.  Also, it has been shown that heavy quarks should be vector-like if the vacuum is to be stable perturbatively\cite{fujikawa}.  The presence of these iso-singlet quarks leads to non-unitarity of the quark mixing matrix and consequently, results in nonvanishing flavour changing neutral currents (FCNC) at the tree level.  On the other hand, the existence of the extra heavy quarks necessitates the modification of the Wilson coefficients of the operators entering the effective Hamiltonian for rare decays of K and B mesons\cite{bvqm}.

In this paper, we investigate the effects of adding an extra iso-singlet pair of quarks, $U$ and $D$ with charges $+2/3$ and $-1/3$ respectively, to the SM on the inclusive rare dileptonic B decays $B\to X_s\ell^+\ell^-$.  Using the experimental upper bound of the branching ratio, one can constrain the acceptable range of the model parameters.  However, it is shown that the shapes of the differential branching ratio and forward-backward asymmetry distribution is still quite sensitive to these parameters in their restricted domain.  We then proceed to calculate the direct CP asymmetry distribution associated with $B\to X_s\ell^+\ell^-$ assuming that the imaginary parts of the $c\bar c$ continuum and resonance contributions are the only sources of the strong phase.  Our results indicate that the size of the CP asymmetry, which is vanishingly small in the SM and very sensitive to the model parameters in VQM, can reach up to 10\% for certain dilepton invariant mass values.
 
\section{Vector-like quark model(VQM)}
In this model, the gauge structure of the SM remains intact except for an additional pair of iso-singlet quarks, which we denote them by $U$ and $D$, as mentioned before.  The difference between these new quarks and ordinary quarks of the three SM generations is that, unlike the latter ones, both left- and right-handed components of the former quarks are $SU(2)_L$ singlets.  Therefore, the Dirac mass terms of vector-like quarks, i.e.
\begin{equation}
m_U(\bar U_L U_R+\bar U_R U_L)+m_D(\bar D_L D_R+\bar D_R D_L) \;\; ,
\end{equation}
are invariant under electroweak gauge symmetry.  However, the masses of the ordinary quarks arise from their gauge invariant Yukawa couplings to an iso-doublet scalar Higgs field $\phi$ as follows:
\begin{equation}
-f^{ij}_d{\bar\psi_L}^id_R^j\phi -f^{ij}_u{\bar\psi_L}^iu_R^j\tilde\phi \; +\; H.C. \;\; ,
\end{equation}
where $i,j=1,2,3$ is covering the three generations of the regular quarks, and the doublet of fermions $\psi_L^i$ is defined as
\begin{equation}
\psi_L^i\equiv {\left (\matrix{
                       u^i \cr
                       d^i} \right )}_L \;\; .
\end{equation}
At the same time, additional $SU(2)_L$ invariant Yukawa couplings between vector-like and ordinary quarks, in the form
\begin{equation}
-f^{i4}_d{\bar\psi_L}^iD_R\phi -f^{i4}_u{\bar\psi_L}^iU_R\tilde\phi \; +\; H.C. \;\; ,
\end{equation}
leads to mixing among 4 up- and down-type quarks of the same charge.  As a result, after spontaneous electroweak symmetry breaking due to $\langle \phi \rangle =v\neq 0$, we obtain the following mass terms:
\begin{equation}
{\bar d_L}^\alpha M_d^{\alpha\beta}d_R^\beta + {\bar u_L}^\alpha M_u^{\alpha\beta}u_R^\beta \; +\; H.C. \;\; ,
\end{equation}
where $M_d$ and $M_u$ being $4\times 4$ mass matrices and $\alpha ,\beta =1..4$ cover ordinary and vector-like quarks.  In general, the mass matrices are not diagonal and unitary transformations from weak to mass eigenstates are necessary to achieve diagonalization.  Denoting the mass eigenstates with $u_{L,R}'$ and $d_{L,R}'$, we have 
\begin{equation}
u^\alpha_{L,R}={A^u_{L,R}}^{\alpha \beta}{u'_{L,R}}^\beta \; ,\;  d^\alpha_{L,R}={A^d_{L,R}}^{\alpha \beta}{d'_{L,R}}^\beta \;\; ,
\end{equation}
where the unitary transformation matrices $A^{u,d}_{L,R}$ are chosen such that ${A^d_L}^\dagger M_d A_R^d$ and ${A^u_L}^\dagger M_u A_R^u$ are diagonal.  The interesting property of the VQM is that the transformations (6) lead to inter-generational mixing among quarks not only in the charged current sector but also in the neutral current interactions.  This is due to the fact that the extra iso-singlet quarks carry zero weak isospin and thus, are not involved in $SU(2)_L$ interactions as weak eigenstates.  For example, the charge current interaction term
\begin{equation}
{J^W_{CC}}^\mu=\sum^3_{i=1}I\frac{g}{\sqrt{2}}{\bar u_L}^i\gamma^\mu d_L^iW^+_\mu \; +\; H.C. \;\; ,
\end{equation}
transforms to 
\begin{equation}
{J^W_{CC}}^\mu=\sum_{\alpha ,\beta =1}^4 I\frac{g}{\sqrt{2}}{\bar {u'}_L}^\alpha V^{\alpha \beta}\gamma^\mu {d'_L}^\beta W^+_\mu \; +\; H.C. \;\; ,
\end{equation}
where
\begin{equation}
V^{\alpha \beta}=\sum_{i=1}^3{({A^u_L}^\dagger)}^{\alpha i}{(A_L^d)}^{i\beta}\;\; ,
\end{equation}
when expressed in terms of mass eigenstates via eqn. (6).  $V$ is the $4\times 4$ generalization of the Cabibbo-Kobayashi-Maskawa(CKM)\cite{ckm} quark mixing matrix.  The fact that the forth generation is iso-singlet, i.e. $i=1,2,3$, leads to non-unitarity of the mixing matrix $V$ as demonstrated in the following:
\begin{eqnarray}
\nonumber {(V^\dagger V)}^{\alpha \beta}&=&\sum_{\delta =1}^4 {V^{\delta\alpha}}^*V^{\delta\beta}=\sum_{\delta =1}^4\sum_{i,j=1}^3{\left [{({A^u_L}^{i\delta})}^* {A^d_L}^{i\alpha}\right ]}^*{({A^u_L}^{j\delta})}^*{A^d_L}^{j\beta} \\
\nonumber &=&\sum_{i,j=1}^3{({A^d_L}^{i\alpha})}^*{A^d_L}^{j\beta}\sum_{\delta =1}^4{({A^u_L}^{j\delta})}^* {A^u_L}^{i\delta} \\
&=&\sum_{i=1}^3{({A^d_L}^{i\alpha})}^*{A^d_L}^{i\beta}=\delta^{\alpha\beta}-{({A_L^d}^{4\alpha})}^*{A_L^d}^{4\beta}\;\; .
\end{eqnarray}
In obtaining the last line in (10), the unitarity of the tranformation matrix $A_L^u$ has been utilized. In the same way, one can show
\begin{equation}
{(VV^\dagger )}^{\alpha\beta}=\delta^{\alpha\beta}-{({A_L^u}^{4\alpha})}^*{A_L^u}^{4\beta}\;\; . 
\end{equation}
Equations (10) and (11), together with the unitarity of $A^{u,d}_L$, indicate that the quark mixing matrix $V$ {\it can not} be unitary.  This property of the VQM has interesting consequences in the neutral current sector where non-vanishing tree level FCNC, proportional to the deviation of the quark mixing matrix (9) from unitarity, are generated.  To demonstrate this explicitly, let us examine the neutral current which is coupled to $Z_\mu$ boson
\begin{equation}
{J^Z_{NC}}^\mu=I\frac{g}{\cos\theta_w}\left (I^q_w\sum_{i=1}^3{\bar q_L}^i\gamma^\mu q^i_L-Q_q\sin^2\theta_w\sum_{\delta=1}^4({\bar q_L}^\delta\gamma^\mu q_L^\delta +{\bar q_R}^\delta\gamma^\mu q_R^\delta )\right ) \;\; ,
\end{equation}
where $Q_q$ is the electric charge of the quark $q$.  The first term in (12) is proportional to $I^q_w$, the third component of the isospin, which has the value $+1/2$ or $-1/2$ for the up- or down-type quarks, respectively.  Consequently, the iso-singlet quarks, which have zero isospin, {\it do not} contribute to this term.  As a result, under the transformations (6), the neutral current (12) can be expressed in terms of the mass eigenstates as follows:
\begin{eqnarray}
\nonumber {J^Z_{NC}}^\mu&=&I\frac{g}{\cos\theta_w}\left (I^q_w\sum_{i=1}^3\sum_{\alpha ,\beta =1}^4{\bar{q'}_L}^\alpha\gamma^\mu {q'_L}^\beta {({A_L^q}^\dagger )}^{\alpha i}{A_L^q}^{i\beta} \right. \\
\nonumber &{}&\left. -Q_q\sin^2\theta_w\sum_{\delta =1}^4\sum_{\alpha ,\beta =1}^4 {\bar{q'}_L}^\alpha\gamma^\mu {q'_L}^\beta {({A_L^q}^\dagger )}^{\alpha \delta}{A_L^q}^{\delta\beta}+{\bar{q'}_R}^\alpha\gamma^\mu {q'_R}^\beta {({A_R^q}^\dagger )}^{\alpha \delta}{A_R^q}^{\delta\beta} \right ) \\
&=&I\frac{g}{\cos\theta_w}\sum_{\alpha ,\beta =1}^4\left (I^q_wU^{\alpha\beta}{\bar{q'}_L}^\alpha\gamma^\mu {q'_L}^\beta -Q_q\sin^2\theta_w\delta^{\alpha\beta}{\bar{q'}}^\alpha\gamma^\mu {q'}^\beta \right ) \;\; ,
\end{eqnarray}
where 
\begin{equation}
U^{\alpha\beta}=\sum_{i=1}^3{({A_L^q}^{i\alpha})}^*{A_L^q}^{i\beta}=\delta^{\alpha\beta}-{({A_L^q}^{4\alpha})}^*{A_L^q}^{4\beta}=\left \{\matrix{
                       {(V^\dagger V)}^{\alpha\beta}\; ,\; q\equiv {\rm down-type} \cr
                       {(V V^\dagger)}^{\alpha\beta}\; ,\; q\equiv {\rm up-type}} \right.\;\; .
\end{equation}
We observe that the non-unitarity of the mixing matrix $V$ in the VQM leads to the tree level FCNC in the $Z$ sector.  In fact, this property is valid for the neutral currents involving the scalar partner of $Z$ and the Higgs bozon as well.  On the other hand, the same non-unitarity parameters $U^{\alpha\beta}(\alpha\neq\beta$) of eqns. (13) and (14) appear at the one-loop level FCNC as multiplicative factors for terms which are independent of the internal quark mass.  These terms are absent in SM where CKM quark mixing matrix is unitary. 

One can use the existing experimental data to constrain various elements of $U$.  In this work, we are concerned with $U^{sb}$ which appears in rare FCNC $b\to s$ transitions.  In particular, we investigate the shift in the differential branching ratio and total branching ratio, forward-backward and CP asymmetries of the inclusive dileptonic B decays due to the presence of the extra vector-like quarks.  The experimental upper bound on the branching ratio of the $B\to X_s \mu^+\mu^-$ decay is used to constrain $U^{sb}$ and other VQM parameters as is explained in the next section.

\section{Rare B decays $B\to X_s\ell^+\ell^-$ in the VQM}
The low-energy effective Lagrangian for $B\to X_s\ell^+\ell^-$ can be written as follows:
\begin{equation}
L_{\rm eff}=\frac{G_F}{\sqrt{2}}\left (A\bar sL_\mu b\bar\ell L^\mu\ell +B \bar sL_\mu b\bar\ell R^\mu\ell +2m_bC\bar sT_\mu b\bar\ell\gamma^\mu\ell \right )\;\; , 
\end{equation}
where
\begin{equation}
\nonumber L_\mu =\gamma_\mu (1-\gamma_5)\;\; , \;\; R_\mu =\gamma_\mu (1-\gamma_5)\;\; , 
\end{equation}
\begin{equation}
\nonumber T_\mu = i\sigma_{\mu\nu}(1+\gamma_5)q^\nu/q^2 \;\; .
\end{equation}
$m_b$ is the mass of the $b$-quark and $q$ is the total momentum of the $\ell^+\ell^-$ pair.  $A$ and $B$ receive long-distance (LD) contributions from dynamical quark loops and intermediate resonances as well as short-distance (SD) contributions at the tree and one loop level.  $C$ is the coefficient of the SD magnetic moment operator.  In the VQM, the tree level FCNC diagram of fig. 1(a) leads to the following result:
\begin{equation}
A^{\rm SD}_{1(a)}=U^{sb}\left (-1+2\sin^2\theta_W\right )\;\; ,\;\; 
B^{\rm SD}_{1(a)}=U^{sb}\left (2\sin^2\theta_W\right )\;\; ,
\end{equation}
where $\sin^2\theta_W\approx 0.23$ ($\theta_W$ is the weak angle).  On the other hand, the box diagram (fig. 1(b)) contributes to $A^{\rm SD}$ only, as it is purely $(V-A)\otimes (V-A)$:
\begin{equation}
A^{\rm SD}_{1(b)}=-\frac{\alpha}{\pi\sin^2\theta_W}\sum_{\beta =1}^4V_{\beta s}^*V_{\beta b}B_0(x_\beta )\;\; .
\end{equation}
The effective vertex function\footnote{We use the notation which is used in Reference \cite{buras} for the effective box and penguin vertices.} $B_0(x_\beta )$ (subscript "0" indicates that QCD corrections are not included), where $x_\beta =m_\beta^2/M_W^2$ with $m_\beta$ being the mass of the virtual quark in the loop, has the following expression \cite{inamilim}:
\begin{equation}
B_0(x)=\frac{1}{4}\left [ 1+\frac{x}{1-x}+\frac{x\ln x}{{(1-x)}^2}\right ] \;\; .
\end{equation}

In the SM with the CKM quark mixing picture, due to unitarity, the constant term in the effective vertex functions like $B_0(x)$, when summed over all three generations, adds up to zero, and therefore, is usually omitted.  However, in the VQM at one loop order, such constant terms make contributions proportional to the non-unitarity parameter $U^{sb}$, and thus, their significance depends on the specific process which is under investigation.  For example, in the radiative $B\to X_s\gamma$ decay process, where a tree level FCNC like fig. 1(a) is absent, the above mentioned non-unitarity contributions are quite crucial \cite{changkeung}.  For the dileptonic $B\to X_s\ell^+\ell^-$ process however, the dominant non-unitarity $U^{sb}$ effect appears at the tree level (eq. (18)), compared to which the above mentioned contributions are suppressed by a factor $\alpha/\pi$, and thus, can be safely ignored at the leading order.

Photon- and Z-penguin diagrams (figs. 1(c) and 1(d)) also contribute to $A$ and $B$.  Figure 1(c) consists of the usual SM diagrams due to W and its unphysical Higgs partner exchange that now include an extra virtual up-type quark insertion in the loop as well.  On the other hand, fig. 1(d) illustrates penguin diagrams via Z, its unphysical scalar Higgs partner and physical Higgs exchange, which are peculiar to the VQM.  Due to the property $\sum_{\delta =1}^4 U^{\alpha\delta}U^{\delta\beta}=U^{\alpha\beta}$, these latter diagrams can all be shown to be proportional to $U^{sb}$, and therefore, are subleading as compared to the tree level FCNC contribution.  Consequently, one has:
\begin{eqnarray}
\nonumber A^{SD}_{1(c)}&=&-\frac{\alpha}{\pi}\sum_{\beta =1}^4V^*_{\beta s}V_{\beta b}\left [\frac{1}{4}D_0(x_\beta )+(1-\frac{1}{2\sin^2\theta_W})C_0(x_\beta )\right ]\;\; , \\
B^{SD}_{1(c)}&=&-\frac{\alpha}{\pi}\sum_{\beta =1}^4V^*_{\beta s}V_{\beta b}\left [\frac{1}{4}D_0(x_\beta )+C_0(x_\beta )\right ]\;\; ,
\end{eqnarray}
where the effective photon and Z vertex functions are defined as:
\begin{eqnarray}
\nonumber D_0(x)&=&-\frac{4}{9}\ln x +\frac{-19x^3+25x^2}{36{(x-1)}^3} +\frac{x^2(5x^2-2x-6)}{18{(x-1)}^4}\ln x \;\; , \\
C_0(x)&=&\frac{x}{8}\left [\frac{x-6}{x-1}+\frac{3x+2}{{(x-1)}^2}\ln x\right ]\;\; .
\end{eqnarray}

At this point, an explanation is in order.  The Z-penguin diagram in the VQM has a divergent term that is absent in the SM due to the unitarity of the quark mixing matrix which is no longer applicable here.  However, this divergence is removed by renormalizing the tree level FCNC which exists in the VQM Lagrangian.  Therefore, the effective vertex function $C_0$ is renormalization scheme dependent and the expression given in eqn. (22) is in fact, in $\overline {\rm MS}$ scheme. However, due to the reason which was pointed out after eqn. (20), this scheme dependence is not relevant for the dileptonic decay process at hand, where, compared to the tree level, the contribution of a one-loop-level constant term is subleading.

Eqns. (18), (19) and (21) determine the SD part of the coefficients $A$ and $B$:
\begin{eqnarray}
\nonumber A^{SD}&=&A^{SD}_{1(a)}+ A^{SD}_{1(b)}+ A^{SD}_{1(c)}\;\; , \\
B^{SD}&=&B^{SD}_{1(a)}+ B^{SD}_{1(c)}\;\; .
\end{eqnarray}
Coefficient C in eqn. (15) receives SD contributions only and is given as:
\begin{equation}
C=\frac{\alpha}{\pi}\sum_{\beta =1}^4V_{\beta s}^*V_{\beta b}\frac{1}{4}D'_0(x_\beta )\;\; ,
\end{equation}
where the effective magnetic moment vertex function $D'_0$ is evaluated to be:
\begin{equation}
D'_0(x)=-\frac{8x^3+5x^2-7x}{12{(1-x)}^3}+\frac{x^2(2-3x)}{2{(1-x)}^4}\ln x \;\; .
\end{equation}

The LD contributions enter $A$ and $B$ coefficients through charm-quark loop ($c\bar c$ continuum) and the intermediate resonances $\psi$ and $\psi'$:
\begin{equation}
A^{LD}=B^{LD}=\frac{\alpha}{2\pi}V^*_{cs}V_{cb}\left [3C_1+C_2 \right ](\tau^{\rm cont}+\tau^{\rm res})\;\; .
\end{equation}
$C_1$ and $C_2$ are the Wilson coefficients of the current-current operators $O_1$ and $O_2$, respectively, which are defined as follows:
\begin{equation}
O_1=\bar sL^\mu b\bar cL_\mu c\;\; ,\;\; O_2=\bar cL^\mu b\bar sL_\mu c\;\; .
\end{equation}
To avoid the scale dependence issue, which is associated with the QCD corrected Wilson coefficients, we take the combination $3C_1+C_2$ in eqn. (26) as a phenomenological parameter whose magnitude $\vert 3C_1+C_2\vert =0.72$ can be determined from the data on the semi-inclusive $B\to X_s\psi$\cite{dht}.  The $c\bar c$ continuum contribution is obtained from joining the $c$ and $\bar c$ legs of the four-Fermi operators in (27)\cite{gsw}:
\begin{equation}
\tau^{\rm cont}=-g(\frac{m_c}{m_b},z)\;\; ,
\end{equation}
where $z=q^2/m_b^2$ ($q$ the total momentum of the dileptons) and 
\begin{equation}
g(y,z)= \displaystyle\left \{ \matrix{\displaystyle\frac{4}{9}\ln{y^2}-\frac{8}{27}-\frac{16y^2}{9z}+\frac{2}{9}\sqrt{1-\frac{4y^2}{z}}\left (2+\frac{4y^2}{z}\right )\left ( \ln{\frac{\vert 1+\sqrt{1-\frac{4y^2}{z}}\vert}{\vert 1-\sqrt{1-\frac{4y^2}{z}}\vert}}\right )\;\; , \;\; z\ge 4y^2\;\; ,\cr
\displaystyle\frac{4}{9}\ln{y^2}-\frac{8}{27}-\frac{16y^2}{9z}+\frac{4}{9}\sqrt{\frac{4y^2}{z}-1}\left (2+\frac{4y^2}{z}\right )arctan\frac{1}{\sqrt{\frac{4y^2}{z}-1}}\;\; , \;\; z\le 4y^2\;\; .}
\right. 
\end{equation}
On the other hand, the resonance contributions from $\psi$ and $\psi'$ can be incorporated by using a Breit-Wigner form for the resonance propagator\cite{resonance}:
\begin{equation}
\tau^{\rm res}=\displaystyle \frac{16\pi^2}{9}\left (\frac{f_\psi^2(q^2)/m_\psi^2}{m_\psi^2-q^2-im_\psi\Gamma_\psi}+(\psi\to\psi')\right )\;\; .
\end{equation}
$f_{\psi (\psi' )}$ is the decay constant of the vector meson, which is defined as:
\begin{equation}
\langle 0\vert \bar c\gamma_\mu c\vert \psi (\psi')\rangle =f_{\psi (\psi')}\epsilon_\mu \;\; ,
\end{equation}
where $\epsilon_\mu$ is the polarization vector. $m_{\psi (\psi')}$ and $\Gamma_{\psi (\psi')}$ are the mass and the total decay width of $\psi (\psi')$, respectively.  In fact, the total branching ratio of $B\to X_s\ell^+\ell^-$ is dominated by these two resonances.  However, by using some appropriate cuts around the resonances in the differential branching ratio, one can get information on the contributing SD physics without any significant interference from the LD sector\cite{ahmady96}.  The fact that there are various observable distributions associated with the dileptonic rare B decays makes these processes excellent venues for examining relevant SD operators within the SM and beyond.

\section{Results and discussion}
By inserting the coefficients $A=A^{\rm SD}+A^{\rm LD}$, $B=B^{\rm SD}+B^{\rm LD}$ and $C$, which are obtained from eqns. (23), (24) and (26), into the effective Lagrangian (15), one can calculate various observables of the decay mode $B\to X_s\ell^+\ell^-$.  The differential decay rate for this process, approximated as the free quark decay $b\to s\ell^+\ell^-$, can be written as:
\begin{eqnarray}
\nonumber\frac{1}{BR(B\to X_c e\bar{\nu_e})}\frac{d BR(B\to X_s\ell^+\ell^-)}{dz}&=&\frac{2{(1-z)}^2}{f(m_c/m_b){\vert V_{cb}\vert}^2} \\
\nonumber &\times & \left (({\vert A\vert}^2+{\vert B\vert}^2)(1+2z)+2{\vert C\vert}^2(1+2/z)+6\Re{\left [{(A+B)}^*C\right ]}\right )\;\; , \\
\end{eqnarray}
where 
\begin{equation}
\nonumber f(x)=1-8x^2+8x^6-x^8-24x^4\ln x\;\; .
\end{equation}
By normalizing to the semileptonic rate in (32), the strong dependence on the b-quark mass is canceled out.

The VQM parameters appearing in (32) are the following: The U-quark mass $m_U$, the non-unitarity parameter $U^{sb}=\vert U^{sb}\vert e^{i\theta}$, where $\theta$ is a weak phase, and $V^*_{4s}V_{4b}=U^{sb}-{(V^\dagger_{\rm CKM}V_{\rm CKM})}^{sb}$.  $V_{\rm CKM}$, which is our notation for the $3\times 3$ submatrix of the matrix $V$, consists of the elements representing mixing among three ordinary generations of quarks.  Therefore, ${(V_{\rm CKM}^\dagger V_{\rm CKM})}^{sb}$, which is zero in the SM, gives a measure of the deviation from unitarity of the 3-generation CKM mixing matrix in the VQM context.

Using the experimental upper bound $BR(B\to X_s\mu^+\mu^-)\le 5.8\times 10^{-5}$\cite{exp}, and assuming the dominance of the tree level contribution (fig. 1(a)), one can extract the following rough constraint on the magnitude of $U^{sb}$:
\begin{equation}
\vert U^{sb}\vert\stackrel{<}{\sim}10^{-3}
\end{equation}
In this work, by taking the above upper limit as a guide, we include all leading contributing factors to the dileptonic rare B decay process in obtaining constraints on the VQM model parameters.  As we pointed out before, due to the tree level $U^{sb}$ contribution to the above decay channels, the terms proportional to $V^*_{4s}V_{4b}$, which appear at the one-loop order, are significant only if $V^*_{4s}V_{4b}\approx -{(V^\dagger_{\rm CKM}V_{\rm CKM})}^{sb}\gg U^{sb}$.  Therefore, having this condition in mind, we parametrize our results in terms of ${(V^\dagger_{\rm CKM}V_{\rm CKM})}^{sb}/\vert V_{cb}\vert =\epsilon e^{i\phi}$ instead of $V^*_{4s}V_{4b}$, where $\epsilon =\vert {(V^\dagger_{\rm CKM}V_{\rm CKM})}^{sb}\vert /\vert V_{cb}\vert$ and $\phi$ is another weak phase of the model. 

As for the numerical values of the ordinary CKM matrix elements, we use $\vert V_{cs}\vert\approx 0.97$, $\vert V_{cb}\vert\approx 0.04$ and $\vert V_{ts}\vert/\vert V_{cb}\vert\approx 1.1$\cite{pdg}, which are extracted from various experimental measurements and are not affected by the presence of the new physics.  We take $V_{us}^*V_{ub}\approx 0$ and assume $V_{cs}^*V_{cb}$ to be real, as is the case, to a good accuracy, in the "standard" parametrization of the CKM matrix.  As a result, $V_{ts}^*V_{tb}$, which is not known experimentally,  can be expressed in terms of the VQM parameters as:
\begin{equation}
\frac{V_{ts}^*V_{tb}}{\vert V_{cb}\vert}\approx \epsilon e^{i\phi}-\vert V_{cs}\vert \;\; .
\end{equation}

In figure 2, the differential branching ratio (32) for some values of the VQM parameters is campared to the SM prediction.  To illustrate the effect of various interfering factors, we use $\epsilon =0.3$ and all constructive/destructive contribution possibilities of the extra beyond the SM terms\footnote{In view of (34), this is more or less the minimum requirement to satisfy the condition $-{(V^\dagger_{\rm CKM}V_{\rm CKM})}^{sb}\gg U^{sb}$}.  We observe that away from the resonances, where SD operators are dominant, the shift from the SM expectation, depending on the parameter values, can be quite significant.  For example, if $\theta$ and $\phi$ are both zero, fig. 2(a) indicates a significant sensitivity of the differential branching ratio to both the non-unitarity parameter $U^{sb}$ and the vector-like up quark mass $m_U$.  Figures 2(b), 2(c) and 2(d) illustrate a reduced sensitivity to these parameters when the new terms generated by the additional quarks enter the effective Lagrangian with a non-vanishing phase.  At the same time, the variation of the differential branching ratio with the invariant mass of the dileptons is appreciably larger than that of the pure SM graph and follows the above-mentioned trend with respect to the relative phase of the extra contributions.

To constrain the model parameters by using the experimental results on $BR(B\to X_s\mu^+\mu^-)$ reported in Ref. \cite{exp}, we calculate the total branching ratio by integrating (23) over all the available range of the dileptonic invariant mass but excluding the resonances $\psi$ and $\psi'$ with a $\delta =\pm 0.1{\rm GeV}$ cut.  Our results are depicted in fig. 3 in the form of  acceptable regions in the $\vert U^{sb}\vert$ versus $m_U$ plane for various choices of the relative sign of the extra contributions.  As is expected, the most stringent constraint is obtained if the relative phases $\theta$ and $\phi$ both vanish (fig. 3(a)).  In this case, absolute upper bounds $\vert U^{sb}\vert\le 1.5\times 10^{-3}$ and $m_U\le 750{\rm GeV}$ can be inferred for these model parameters.  These limits become stronger if a larger value for $\epsilon$ is used.  Figures 3(b) and 3(c) show that, if the extra tree and penguin contributions have positive relative sign, the experimental bound leads to a lower limit for $m_U$ for larger values of the non-unitarity parameter $U^{sb}$.  A less stronger upper bounds on these parameters are resulted when $\theta$ and $\phi$ are both non-zero (fig. 3(d)).

Besides the differential and total branching ratios,  there are other physical observables associated with the dileptonic rare B decays which can provide crucial information on the contributing amplitudes and their relative phases.  Among them are the forward-backward and the CP asymmetry distributions, which we investigate in the presence of the extra vector-like quarks.  The forward-backward asymmetry distribution is defined as:
\begin{equation}
A_{FB}(z)=\frac{\int_0^1dwd^2BR/dwdz-\int_{-1}^0dwd^2BR/dwdz}{\int_0^1dwd^2BR/dwdz+\int_{-1}^0dwd^2BR/dwdz}\;\; ,
\end{equation}
where $w=\cos\omega$, with $\omega$ being the angle between the momentum of the ingoing B meson (or the outgoing s quark) and that of $\ell^+$ in the center of mass frame of the dileptons.  Using the effective Lagrangian (15), a simple form for this asymmetry in the $m_s=0$ limit is obtained\cite{amm}:
\begin{equation}
A_{FB}(z)=\frac{3}{2}\frac{\left ({\vert A\vert}^2-{\vert B\vert}^2\right )z+2\Re{\left [(A+B)^*C\right ]}}{\left ({\vert A\vert}^2+{\vert B\vert}^2\right )(1+2z)+2{\vert C\vert}^2(1+2/z)+6\Re{\left [(A+B)^*C\right ]}}\;\; .
\end{equation}
Figure 4 illustrates our results for the forward-backward asymmetry distribution of the decay $B\to X_s\mu^+\mu^-$ in the VQM as compared to the SM.  From fig. 4(a) and 4(b) we observe that, when the tree level FCNC contribute constructively, even though the sign of the asymmetry remains the same as in the SM, its shape, away from the resonances, can be significantly different.  On the other hand, as is shown in figs. 4(c) and 4(d), for the destructive contribution of the non-unitarity induced tree level term and large enough values of $\vert U^{sb}\vert$ or $m_U$, the sign of $A_{FB}$ can be opposite to what is predicted by the SM.  One important observable in this decay channel is the point of zero asymmetry in the forward-backward asymmetry distribution, which occurs somewhere below the resonance $\psi$.  Our investigation reveals that, as far as the VQM is concered, the position of this point is quite stable and is not shifted very much from its SM value for various choices of the model parameters.

The CP asymmetry distribution is defined as:
\begin{equation}
A_{CP}(z)=\displaystyle\frac{\frac{dBR}{dz}(B\to X_s\ell^+\ell^-)-\frac{dBR}{dz}(\bar B\to \bar X_s\ell^+\ell^-)}{\frac{dBR}{dz}(B\to X_s\ell^+\ell^-)+\frac{dBR}{dz}(\bar B\to \bar X_s\ell^+\ell^-)}\;\; ,
\end{equation}
where the decay rate for the charge conjugate process $\bar B\to\bar X_s\ell^+\ell^-$ is obtained from the rate for $B\to X_s\ell^+\ell^-$ by reversing the sign of the weak phases $\theta$ and $\phi$.  Since in the VQM, the effective Lagrangian (15) contains terms with different CP-odd weak phases, as well as, LD continuum and resonance contributions which are sources of perturbative and nonperturbative CP-even strong phases, the direct CP asymmetry (38), unlike the SM, is expected to be non-zero.  The size of this asymmetry, which is zero for invariant dilepton masses below $2m_c$ threshold due to the vanishing strong phase, depends on the interplay of the various contributing terms.  Our results for certain choices of the model parameters are depicted in fig. 5.  From fig. 5(a) and 5(b), we observe that when the weak phase $\phi$ is large, smaller values of the non-unitarity parameter lead to significantly larger CP asymmetries sensitive to the $U$ quark mass, $m_U$.  In fact, for a negative relative sign of the tree level contribution ($\theta =\pi$) and $m_U=400 {\rm GeV}$, as is shown in fig. 5(b), asymmetries of the order of 10\% can be achieved.  On the other hand, for smaller values of $\phi$, larger $\vert U^{sb}\vert$ values can generate asymmetries of the order of a few percent if $m_U\ge 400{\rm GeV}$(fig. 5(c)).  Finally, we find out that for a purely real deviation of $V_{\rm CKM}$ from unitarity, i.e. $\phi =0$ or $\pi$, the CP asymmetries away from the resonances are smaller.   Figure 5(d) shows a particular instance of this situation where off-resonance-peak CP asymmetries are less than 2\%, at most.

In conclusion, we have investigated various observables of the dileptonic rare B decay $B\to X_s\ell^+\ell^-$ in the presence of an extra generation of vector-like quarks.  The measurement of this decay channel in the near future should provide more stringent constraints on the model parameters.  We showed that the shape of the differential branching ratio and forward-backward asymmetry distribution can be quite distinct from the SM predictions.  CP asymmetry in this decay mode, which is near zero in the SM, shows significant sensitivity to the VQM parameters.  Asymmetries of up to 10\% can be achieved for a typical choice of the parameters which is used in our investigation.

\section{acknowledgement}
M. A. acknowledges support from the Japanese Society for the Promotion of Science under a JSPS long-term visiting fellowship.

\end{document}